\documentclass[reqno]{amsart}

\usepackage{booktabs}
\usepackage[table,xcdraw]{xcolor}
\usepackage{rotating,tabularx,siunitx}
\newcolumntype{T}{>{\raggedleft\arraybackslash} X}

\usepackage{epstopdf}

\usepackage{graphicx}
\usepackage{enumerate}
\usepackage[implicit=false]{hyperref}
\usepackage{geometry}  % set page parameters suitably
\usepackage{mathtools} % for \smashoperator macro
\usepackage{tikz}

\usepackage[ruled,linesnumbered]{algorithm2e}
\SetKwInput{KwData}{Input}
\SetKwInput{KwResult}{Output}

\usepackage{amssymb,latexsym,amsfonts,amsmath,mathrsfs}

\usepackage{amsthm}

\def\reals{\mathbb{R}}
\def\nats{\mathbb {N}}

\newcommand{\initial}{X_\mathcal{I}}
\newcommand{\unsafe}{X_\mathcal{U}}

\newtheorem{lemma}{Lemma}
\newtheorem{definition}{Definition}
\newtheorem{remark}{Remark}

\def\barrier{\mathcal{B}}

\usepackage{listings}
\usepackage{xcolor}
\usepackage{enumerate}
\definecolor{codegreen}{rgb}{0,0.6,0}
\definecolor{codegray}{rgb}{0.5,0.5,0.5}
\definecolor{codepurple}{rgb}{0.58,0,0.82}
\definecolor{backcolour}{rgb}{0.95,0.95,0.92}
\lstdefinestyle{mystyle}{
	backgroundcolor=\color{backcolour},   
	commentstyle=\color{codegreen},
	keywordstyle=\color{magenta},
	numberstyle=\tiny\color{codegray},
	stringstyle=\color{codepurple},
	basicstyle=\ttfamily\footnotesize,
	breakatwhitespace=false,         
	breaklines=true,                 
	captionpos=b,                    
	keepspaces=true,                 
	numbers=left,                    
	numbersep=5pt,                  
	showspaces=false,                
	showstringspaces=false,
	showtabs=false,                  
	tabsize=2
}
\usepackage{fancyhdr}

\newenvironment{nouppercase}{%
	\renewcommand{\uppercasenonmath}[1]{}}{}

\begin{document}

\begin{abstract}
We develop an open-source software tool, called \textsf{PRoTECT}, for the parallelized construction of safety barrier certificates (BCs) for nonlinear polynomial systems. This tool employs \emph{sum-of-squares (SOS) optimization} programs to systematically search for polynomial-type BCs, while aiming to verify safety properties over \emph{four classes of dynamical systems}: (i) discrete-time \emph{stochastic} systems, (ii) discrete-time \emph{deterministic} systems, (iii)  \emph{continuous-time} stochastic systems, and (iv) \emph{continuous-time} deterministic systems. In particular, \textsf{PRoTECT} is the first software tool that offers stochastic barrier certificates. \textsf{PRoTECT} is implemented in Python as an application programming interface (API), offering users the flexibility to interact either through its user-friendly graphic user interface (GUI) or via function calls from other Python programs. \textsf{PRoTECT} leverages \emph{parallelism} across different barrier degrees to efficiently search for a feasible BC.\\

\end{abstract}

\title{\LARGE \textsf{\bf{PRoTECT}}: \underline{\textbf{P}}arallelized Const\underline{\textbf{R}}uction \underline{\textbf{o}}f Safe\underline{\textbf{T}}y 
Barri\underline{\textbf{E}}r 
 \underline{\textbf{C}}ertificates for Nonlinear Polynomial Sys\underline{\textbf{T}}ems}

\author{{\bf {\large Ben Wooding, Viacheslav Horbanov, and Abolfazl Lavaei}}\\{\normalfont School of Computing, Newcastle University, United Kingdom}\\ \textsf{\{ben.wooding,v.horbanov2,abolfazl.lavaei\}@newcastle.ac.uk}}

\pagestyle{fancy}
\lhead{}
\rhead{}
  \fancyhead[OL]{Ben Wooding, Viacheslav Horbanov, Abolfazl Lavaei}

  \fancyhead[EL]{\textbf{\textsf{PRoTECT}}: Parallelized Construction of Safety Barrier Certificates for Polynomial Systems}
  \rhead{\thepage}
 \cfoot{}
 
\begin{nouppercase}
	\maketitle
\end{nouppercase}

\section{Introduction}
\subsubsection{Motivation for \textsf{PRoTECT}.}

Formal verification of dynamical systems has become a focal point over the past several years, primarily due to their widespread integration into safety-critical systems~\cite{knight2002safety}. These systems, whose malfunction may lead to severe consequences such as loss of life, injuries, or substantial financial losses, are integral across a broad spectrum of domains including robotics, transportation systems, energy, and healthcare. When confronted with a property of interest for a dynamical model, formal verification aims to reliably assess whether the desired specification is fulfilled. If the model exhibits stochastic behavior, the objective shifts to formally quantifying the satisfaction probability of the desired property.

Barrier certificates (BCs), also known as \emph{barrier functions}, have emerged as a fundamental solution approach, offering assurances regarding the safety behavior of diverse classes of systems. Specifically, BCs can be employed to directly assess the behavior of systems across \emph{continuous-state spaces} with an uncountable number of states, without resorting to discretization, which contrasts with abstraction-based approaches~\cite{lavaei2022survey}. This aspect is particularly noteworthy when considering \emph{safety},  (\emph{a.k.a. invariance})  properties, wherein the state transitions of the system remain within a region labeled as ``safe'', ensuring no transitions occur to any region labeled ``unsafe''. In particular, barrier certificates, akin to Lyapunov functions, are functions established over the system's state space, fulfilling specific inequalities concerning both the function itself and the one-step transition (or the flow) of the system. A suitable level set of a BC can segregate an unsafe region from all system trajectories originating from a specified set of initial conditions. Hence, the presence of such a function offers a formal (probabilistic) certification for system safety.

\subsubsection{Related Literature on BCs.}

Barrier certificates, initially introduced in~\cite{prajna2004stochastic,prajna2007framework}, have gained significant recognition for the formal safety analysis of dynamical systems over the past 20 years. These approaches can verify systems with polynomial dynamics, enabling the design of polynomial BCs using sum-of-squares techniques, \emph{e.g.,} via SOSTOOLS~\cite{SOSTOOLS}. Additionally, some alternative approaches  explore verifying nonpolynomial systems, such as those discovered through counter-example guided inductive synthesis (CEGIS), leveraging satisfiability modulo theories (SMT) solvers including Z3 \cite{de2008z3}, dReal \cite{gao_-complete_2012}, or MathSat \cite{cimatti_mathsat5_2013}. Other new techniques encompass neural barrier functions~\cite{zhao2020synthesizing,mathiesen2022safety} and genetic programs~\cite{verdier2018formal}. A comprehensive overview of barrier certificates  can be found in~\cite{ames2019control,xiao2023safe}. While we primarily focus on utilizing BCs for \emph{safety} specifications in \textsf{PRoTECT}, they also hold significant value for addressing other specifications, including reachability, reach-while-avoid, and other temporal logic specifications~\cite{anand2024compositional,Lindemann2020CBF4multirobotSTL}.

\subsubsection{Related Software Tools.}
\textsf{FOSSIL}~\cite{abate2021fossil} is a relevant tool for the \emph{model-based} construction of barrier certificates. In the latest release~\cite{edwards2023fossil}, \textsf{FOSSIL} has expanded its capabilities to include finding barrier certificates for discrete- and continuous-time \emph{deterministic} systems using the CEGIS approach, while facilitating verification and control synthesis for specifications including safety, reachability, and reach-while-avoid. However, \textsf{FOSSIL} lacks support for \emph{stochastic} systems, whereas in \textsf{PRoTECT}, we provide support for both discrete- and continuous-time \emph{stochastic} systems. Additionally, while \textsf{FOSSIL} can handle nonpolynomial BCs, it does not guarantee termination due to its use of CEGIS approach. In contrast, \textsf{PRoTECT} utilizes sum-of-squares optimization techniques, enabling the exploitation of \emph{parallelism} to concurrently search for multiple candidate BCs with different degrees, aiming to construct them effectively. Recently, two new tools for constructing barrier certificates have been introduced. \textsf{TRUST} \cite{gardner2025trust} is a \emph{data-driven} tool that generates barrier certificates for deterministic systems with \emph{unknown} polynomial dynamics, using only a single trajectory of collected data. In addition, \textsf{CBFKIT} \cite{black2024cbfkit} is a toolbox designed for safe robotic planning. It supports both deterministic and stochastic continuous-time dynamics but requires the user to provide a barrier function \textit{a priori}, which it then verifies for correctness—unlike \textsf{PRoTECT}, which automatically synthesizes a barrier certificate to meet the required conditions.

\subsubsection{Original Contributions.}
The primary contributions and noteworthy aspects of our tool paper are as follows:

\begin{enumerate}[(i)]
    \item We propose the first tool, employing  SOS optimization techniques, that verifies the safe behavior of \emph{four classes of dynamical systems}: (i) discrete-time \emph{stochastic} systems (dt-SS), (ii) discrete-time \emph{deterministic} systems (dt-DS), (iii)  \emph{continuous-time} stochastic systems (ct-SS), and (iv) \emph{continuous-time} deterministic systems (ct-DS). In particular, \textsf{PRoTECT} is the first software tool that offers stochastic barrier certificates.
    \item \textsf{PRoTECT} is implemented in Python using \textsf{SumOfSquares}~\cite{Yuan_SumOfSquares_py}, and leverages \emph{parallelization} to efficiently search for BCs of different degrees, aiming to satisfy the desired safety specifications.
    \item \textsf{PRoTECT} supports \emph{normal}, \emph{uniform}, and \emph{exponential} noise distributions for dt-SS, as well as \emph{Brownian motion} and \emph{Poisson processes} for ct-SS.
    \item \textsf{PRoTECT} offers advanced GUIs for all four classes of models, enhancing the tool's accessibility and user-friendliness.
    \item We utilize \textsf{PRoTECT} across various real-world applications, including room temperature systems, a jet engine, a DC motor, a Van der Pol oscillator, a model of two fluid tanks, and an $8$-dimensional system to showcase the scalability. This expansion broadens the applicability of formal method techniques to encompass safety-critical applications across different model classes. The results highlight significant improvements in computational efficiency.
\end{enumerate}

The source code for \textsf{PRoTECT}, along with detailed guidelines on installation and usage, including tutorial videos, are available at:
\begin{center}
	\href{https://github.com/Kiguli/ProTECT}{https://github.com/Kiguli/PRoTECT}
\end{center}

\section{Problem Description}

\subsubsection{Preliminaries and Notation.}
The set of real numbers, non-negative and positive real numbers is denoted by $\reals$, $\reals^+_0$, and $\reals^+$, respectively. The set of natural numbers including and excluding zero is represented by $\nats$ and $\nats^+$. The empty set is denoted by $\emptyset$. For any matrix $A \in \mathbb R^{n\times n}$,  $\mathbf{Tr}(A)$ represents the trace of $A$ which is the sum of all its diagonal elements.  For a system $\Sigma$ and a property $\phi$, the notation $\Sigma\models \phi$ signifies that $\Sigma$ fulfills the property $\phi$. We consider a probability space ($\Omega, \mathcal{F}_\Omega, \mathbb{P}_\Omega$), where $\Omega$ represents the sample space, $\mathcal{F}_\Omega$ denotes the sigma-algebra of $\Omega$ comprising events, and $\mathbb{P}_\Omega$ signifies the probability measure assigning probabilities to each event. A topological space $S$ is termed a Borel space when equipped with a metric that renders it a separable and complete metrizable space. Subsequently, $S$ is furnished with a Borel sigma-algebra denoted as $\mathscr{B}(S)$. For continuous-time stochastic systems, we assume that triple ($\Omega, \mathcal{F}_\Omega, \mathbb{P}_\Omega$) is
equipped with a filtration $\mathbb F = (\mathcal{F}_s)_{s\geq0}$ adhering to the standard conditions of completeness and right continuity. Consider $(\mathbb W_s)_{s\geq0}$ as a $b$-dimensional $\mathbb F$-Brownian motion, and $(\mathbb{P}s)_{s\geq0}$ as an $r$-dimensional $\mathbb F$-Poisson process. We posit independence between the Poisson process and Brownian motion. Poisson
process $\mathbb P_s = [\mathbb P_s^1;\ldots; \mathbb P_s^r]$ represents $r$ events, each assumed to occur independently of the others.
\begin{figure}[t]
	\centering
		\resizebox{0.5\linewidth}{!}{\begin{tikzpicture}
		% Large Oval
		\node at (2,2) {$X$};
		\draw[thick] (0,0) ellipse (4.5cm and 2.5cm);
		% First smaller Oval with dashed barrier
		\draw[thick,fill=blue!20, blue,opacity=0.4] (-1.5,0) ellipse (1.5cm and 1.5cm);
		\node at (-1.5,1) {$X_\mathcal{I}$};
		\node at (-1.5,0.5) {$\barrier(x) \leq \gamma$};
		\draw[thick,fill=red!20, red,opacity=0.4] (2.7,0) ellipse (1cm and 1cm);
		\node at (2.7,0.5) {$X_\mathcal{U}$};
		\node at (2.7,0) {$\barrier(x) \geq \lambda$};
		\draw[thick,dashed] (-1,0) ellipse (2.5cm and 2cm);
		% Dashed barrier label
		\node at (0,1.3) {$\lambda > \gamma$};
            \newcommand{\markpoint}[2]{\draw[fill,black] (#1,#2) circle (2pt); \draw[black] (#1-0.15,#2-0.15) -- (#1+0.15,#2+0.15); \draw[black] (#1-0.15,#2+0.15) -- (#1+0.15,#2-0.15);}
    % Mark a point with a cross at (0,0)
    \markpoint{-1.5}{0}
    % Define the snake path
    \draw[thick, black] plot[smooth] coordinates {(-1.5,0) (-1,0) (-1,-0.5) (-0.5,-0.5) (-0.5,-1) (-1,-1) (-1.5,-1.45) (-2,-1) (-2.5,-1) (-2,-0.5) (-1.5,-1)};
	\end{tikzpicture}}
	\caption{A barrier certificate $\barrier(x)$ for a dynamical system. The dashed line denotes the initial level set $\barrier(x) = \gamma$.}
	\label{fig:BC-demo}
\end{figure}

\subsubsection{Safety Barrier Certificates.}

Consider a state set $X$ in an $n$-dimensional space, denoted as $X\subseteq\mathbb{R}^n$. Within this set, we identify two specific subsets: $\initial$ and $\unsafe$, which represent the \emph{initial and unsafe} sets, respectively. The primary objective is to construct a function $\barrier(x)$, termed the \emph{barrier certificate}, along with constants $\gamma$ and $\lambda$ as the \emph{initial} and  \emph{unsafe} level sets of $\barrier(x)$, as illustrated in Fig.~\ref{fig:BC-demo}. Specifically, the design of BC incorporates two conditions concerning these level sets, in conjunction with a third criterion that captures the \emph{state evolution} of the system. Collectively, satisfaction of conditions provides a (probabilistic) guarantee that the system's trajectories, originating from any initial condition $x_0 \in \initial$, will not transition into the unsafe region $\unsafe$.

We now formally introduce the safety specification that we aim to investigate in this work.

\begin{definition}[Safety]\label{Safety}
	A safety specification is defined as $\varphi = (\initial,\unsafe,\mathcal{T})$, where $\initial,\unsafe\subseteq X$ with $\initial\cap\unsafe=\emptyset$, and $\mathcal{T}\in\nats\cup\{\infty\}$. A dynamical system $\Sigma$ is considered safe over an (in)finite time horizon $\mathcal{T}$, denoted as $\Sigma\models_\mathcal{T}\varphi$ if all trajectories of $\Sigma$ starting from the initial set $\initial$ never reach the unsafe set $\unsafe$. If trajectories are probabilistic, the primary goal is to
	compute $\mathbb P \{\Sigma\models_\mathcal{T}\varphi\}\ge \phi$, with $\phi \in[0,1]$.
\end{definition}

\subsubsection{Overview of \textsf{PRoTECT}.}
\textsf{PRoTECT} offers functionalities that automatically generate BCs and verify the safety property across \emph{four distinct classes of systems}. The description of the system serves as an input to the tool, triggering the appropriate function. These functions are named as \verb$dt-SS$, \verb$dt-DS$, \verb$ct-SS$ and \verb$ct-DS$. Additionally, the geometric characteristics for sets of interest, which define the safety specification according to Definition~\ref{Safety}, constitute another input to the tool.  While a GUI is designed to enhance the user-friendliness of \textsf{PRoTECT}, the BCs may also be verified via configuration files executed through the command line. As the output, \textsf{PRoTECT} returns BC $\barrier(x)$, level sets $\gamma$ and $\lambda$, and, for stochastic systems, the value $c$ and the confidence level $\phi$ (cf. Section~\ref{Stochastic}).

Utilizing methodologies from the \emph{sum-of-squares} (SOS) domain, facilitated by the \textsf{SumOfSquares} Python toolbox~\cite{Yuan_SumOfSquares_py}, \textsf{PRoTECT} adopts polynomial structures for BCs expressed as $\barrier(x)=\sum_{j=1}^{z}q_j{p}_j(x)$, with basis functions ${p}_j(x)$ that are monomials over $x$, and unknown coefficients $q = [q_0,\ldots,q_z]\in\reals^z$ that need to be designed. \textsf{PRoTECT} leverages parallelization techniques to facilitate the \emph{simultaneous} verification of multiple BCs, differentiating them based on their polynomial degrees, where degrees must be even~\cite{reznick2000some}. In deterministic systems, upon finding a feasible BC, the parallel processing is terminated and the valid BC is returned to the user. Conversely, for stochastic systems, \textsf{PRoTECT} awaits until all potential solutions are fully processed, subsequently selecting and returning the BC that offers the highest probabilistic confidence. This process is detailed in the provided pseudo-code, illustrated in Algorithm~\ref{alg:parallel-BC}.

\begin{algorithm}[t]
	\caption{\emph{Parallel} Construction of BCs}
	\label{alg:parallel-BC}
	\SetAlgoLined
	\KwData{system $\Sigma$, maximum polynomial degree $P$, \emph{required} parameters $K_{req}$, \emph{optional} parameters $K_{opt}$}
	\texttt{temp = []}\; choose function \texttt{func} for $\Sigma$ to identify the class of system\;
	\ForAll{$p\in \{2,{4,} \ldots, P\}$ \textbf{in parallel}}{
		\texttt{barrier} = \texttt{func}$(p,K_{req},K_{opt})$\;
		\If {$\emph{\texttt{barrier}}$ \emph{is SOS} }{
			\texttt{temp.append(barrier)}\;
			\If{\emph{$\Sigma$ is deterministic}}
			{
				\tcp{terminate all parallel processes}
			}
		}
	}
	\tcp{return element with highest confidence in temp}
	$\barrier(x)$ = \texttt{max(temp)}\;
	\KwResult{barrier certificate $\barrier(x)$, level sets $\gamma$, $\lambda$; confidence $\phi$ and constant $c$ (for dt-SS and ct-SS)}
\end{algorithm}

\section{Discrete-Time Stochastic Systems}\label{Stochastic}
In this section, we define the notion of barrier certificates for discrete-time stochastic systems (dt-SS). A dt-SS is a tuple $\Sigma_d^\varsigma = (X,\varsigma,f)$, where: $X\subseteq \reals^n$ is a Borel space as the state set,  $\varsigma$ is a sequence of independent and identically distributed (i.i.d.) random variables from a sample space $\Omega$ to a measurable set $\mathcal{V}_\varsigma$, \emph{i.e.,} $\varsigma:= \{\varsigma(k)\!:\Omega \rightarrow \mathcal{V}_\varsigma,~k\in\nats\},$
and $f\!:X \times \mathcal{V}_\varsigma\rightarrow X$ is a measurable function characterizing the \emph{state evolution} of the system. For a given initial state $x(0)\in X$, the state evolution of $\Sigma_d^\varsigma$ is characterized by
\begin{equation}\label{eq:dt-SS:dynamics}
	\Sigma_d^\varsigma\!:x(k+1) = f(x(k),\varsigma(k)), \quad k\in\nats.
\end{equation}
The stochastic process $x_{x_0}\!\!: \Omega \times \mathbb{N} \rightarrow X$, which fulfills \eqref{eq:dt-SS:dynamics} for any initial state $x_0 \in X$ is referred to as the \emph{solution process} of dt-SS at time $k \in \mathbb{N}$. \textsf{PRoTECT} accommodates \emph{additive noise} types across a range of distributions, including \emph{uniform}, \emph{normal}, and \emph{exponential} distributions. The notion of barrier certificates for dt-SS is provided by the subsequent definition~\cite{prajna2007framework}.

\begin{definition}[BC for dt-SS]
	\label{def:dt-SS-BC}
	Consider the dt-SS $\Sigma_d^\varsigma = (X,\varsigma,f)$ and $\initial,\unsafe\subseteq X$. A function $\barrier:X\rightarrow\reals^{+}_{0}$ is known as the barrier certificate (BC), if there exists constants $\lambda,\gamma,c\in\reals^{+}_{0}$,  with $\lambda > \gamma$, such that
	\begin{align}
		\label{gamma-condition}
		\barrier(x) \leq \gamma, \quad\quad&\forall x\in \initial, \\
		\label{lambda-condition}
		\barrier(x) \geq \lambda, \quad\quad&\forall x\in \unsafe, \\
		\label{dt-SS-condition}
		\mathbb{E}\Big[\barrier(f(x,\varsigma))~\big\vert~x\Big] \leq \barrier(x) + c, \quad\quad&\forall x\in X,
	\end{align}
	where $\mathbb{E}$ denotes the expected value of the system's one-step transition, taken with respect to $\varsigma$.
\end{definition}

We now leverage the BC in Definition~\ref{def:dt-SS-BC} and quantify a lower bound confidence over the safety of dt-SS~\cite{kushner1965stability,prajna2007framework}. This lemma, commonly found in the literature (e.g. \cite{salamati2021data}), provides the safety confidence for stochastic systems. The same confidence formula applies to \emph{continuous-time} stochastic systems in Section~\ref{sec:ct-SS}.

\begin{lemma}[Confidence $\phi$]\label{Lemma1}
	For dt-SS $\Sigma_d^\varsigma$, let there exist a BC as in Definition~\ref{def:dt-SS-BC}. Then the probability that trajectories of dt-SS starting from any initial condition $x_0\in\initial$ will not reach the unsafe region $\unsafe$ within a finite time horizon $k \in [0,\mathcal T]$ is quantified as
	\begin{equation}\label{confidence}
		\phi = \mathbb{P}\Big\{x_{x_0}(k) \notin \unsafe~\text{for all}~ k \in [0,\mathcal T] \,\big|\, x_0 = x(0) \Big\} \geq 1-\frac{\gamma+c\mathcal{T}}{\lambda}.
	\end{equation}
\end{lemma}

Under the assumption that $f$ is a polynomial function of state $x$ and sets $\initial,\unsafe, X$ are semi-algebraic sets, \emph{i.e.} they can be represented by polynomial inequalities, we now describe how to reformulate~\eqref{gamma-condition}-\eqref{dt-SS-condition} as an SOS optimization program to search for a polynomial-type BC~\cite[Prop. 18]{prajna2007framework}.

\begin{lemma}\label{SOS}
	Let sets $\initial,\unsafe, X$ be defined element-wise by vectors of polynomial inequalities $\initial = \{x\in\reals^n~\vert~{g_i}(x)\geq 0\}, \unsafe = \{x\in\reals^n~\vert~{g_u}(x)\geq 0\},$ and $X = \{x\in\reals^n~\vert~g(x)\geq 0\}$. Suppose there exists an SOS polynomial $\barrier(x)$, constants $ \lambda,\gamma, c \in \reals^+_0$, with $\lambda > \gamma$, and vectors of SOS polynomials ${l_i}(x)$, ${l_u}(x)$, and $l(x)$ such that the following expressions are SOS polynomials:
	\begin{align}
		\label{lagrangian multiplier1}
		-\barrier(x) - {l_i}(x){^\top}{g_i}(x) + \gamma, \\
		\label{lagrangian multiplier2}
		\barrier(x) - {l_u}(x){^\top}{g_u}(x) - \lambda, \\
		\label{lagrangian multiplier dt-SS}
		-\mathbb{E}\Big[\barrier(f(x,\varsigma))~\big\vert~x\Big] +\barrier(x) - l(x){^\top}g(x) + c.
	\end{align}
	Then $\barrier(x)$ satisfies conditions~\eqref{gamma-condition}-\eqref{dt-SS-condition} in Definition~\ref{def:dt-SS-BC}.
\end{lemma}

This lemma is well-established in the literature and extends naturally from works such as~\cite{prajna2004safety,prajna2007framework,lavaei2022survey,jagtap2020formal}.

\begin{remark}
	\textsf{PRoTECT} is equipped to accommodate any \emph{arbitrary number} of unsafe regions $X_{\mathcal{U}_i}$, where $i\in\{1,\ldots, m\}$. In such scenarios, condition \eqref{lagrangian multiplier2} should be reiterated and enforced for each distinct unsafe region.
\end{remark}

\subsection{\textsf{PRoTECT} Implementation for dt-SS}\label{parameter}
\label{sec:GUI_description}

\begin{figure}[t]
    \centering
     \includegraphics[width=0.8\textwidth]{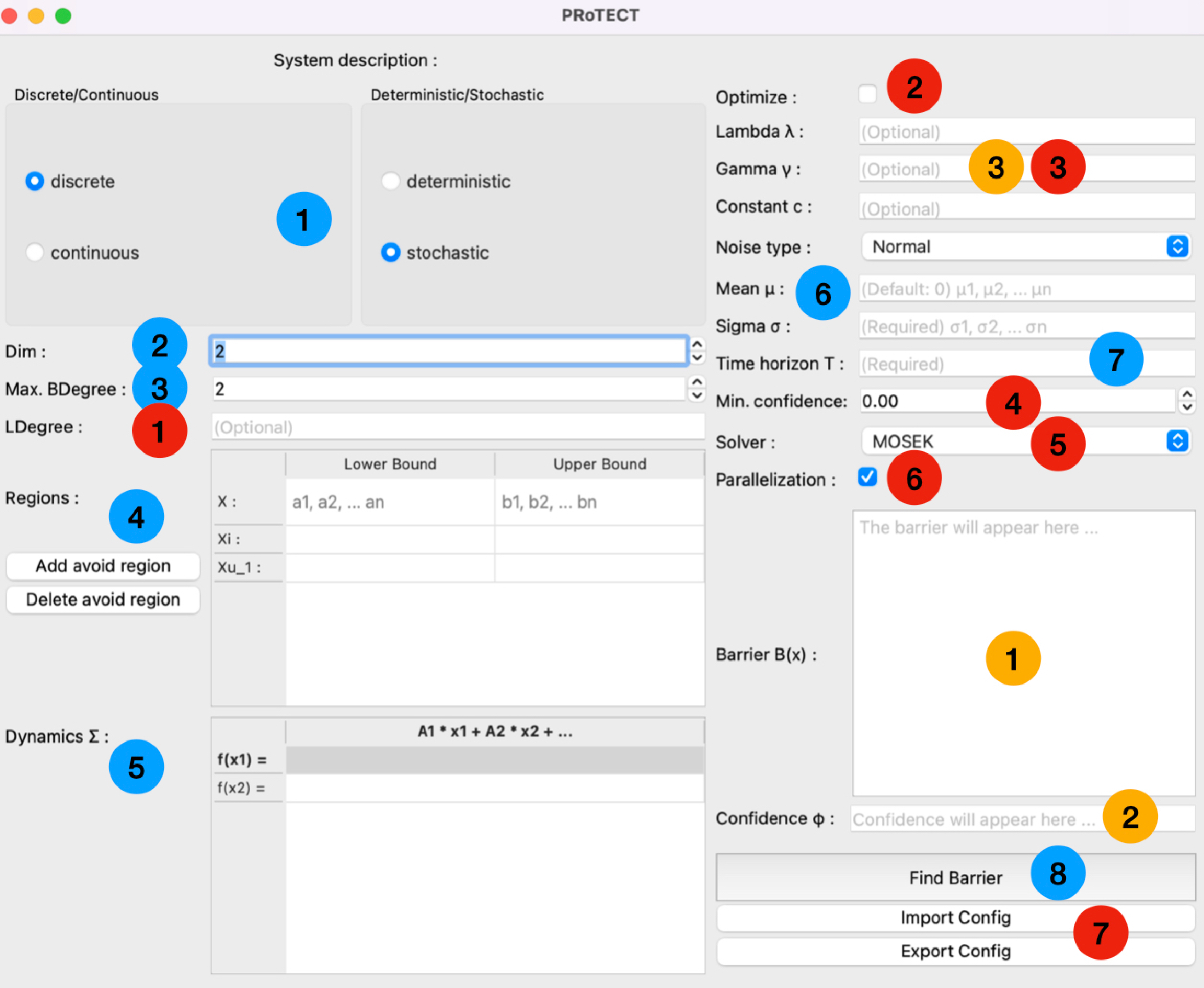}\vspace{-0.2cm}
    \caption{\textsf{PRoTECT} GUI for dt-SS, where required parameters, optional parameters, and outputs are marked with blue, red, and yellow circles, respectively (see Section~\ref{sec:GUI_description}).}
    \label{fig:GUI}
\end{figure}

\subsubsection{Graphic User Interface (GUI).} To enhance accessibility and user-friendliness of the tool, \textsf{PRoTECT} offers the Model-View-Presenter architecture incorporating a GUI. Specifically, a GUI strengthens user-friendliness by abstracting away implementation details for the code, allowing for a push-button method to construct barrier certificates. In Fig.~\ref{fig:GUI}, color notation is utilized to represent labels by their corresponding color and number. While \textsf{PRoTECT} provides GUIs for all four classes of systems (see Fig.~\ref{fig:GUI} (blue-1)), we only depict it for dt-SS due to space constraints.  Our tool offers two implementations, either serial or parallel (red-6). The tool processes the information entered into the GUI before executing the desired function upon pressing the \emph{Find Barrier} button (blue-8). Outputs of barrier certificate $\barrier(x)$, confidence $\phi$, level sets $\gamma$ and $\lambda$, and constant $c$ are displayed at (yellow-1), (yellow-2), and (yellow-3), respectively. Optionally, the GUI allows for the import and export of configuration parameters in JSON format using the \emph{Import Config} and \emph{Export Config} buttons (red-7), with examples available in the folder  \texttt{/ex/GUI\_config\_files}.

\subsubsection{Application Programming Interface (API)} In general, the backend of \textsf{PRoTECT} behaves as an API, with functions that can be called and used in any python program. In the project folders \texttt{/ex/benchmarks-stochastic} and \texttt{/ex/benchmarks-deterministic}, we provide some generic configuration files which demonstrate how to use the functions in a standard python program. The user is expected to provide the following \emph{required} parameters: dimension of the state set $X\subseteq \reals^n$ (blue-2), indicated by \verb|dim|, and the degree of the barrier certificate (blue-3), denoted by \verb|b_degree|. The lower and upper bounds of the initial region $\initial$ , labeled as \verb|L_initial| and \verb|U_initial|; lower and upper bounds of the unsafe region $\unsafe$, referred to as \verb|L_unsafe| and \verb|U_unsafe|; lower and upper bounds for the state set $X$, denoted as \verb|L_space| and \verb|U_space|; where the value of each dimension is separated with a comma (blue-4). Due to possible scenarios with multiple unsafe regions, the unsafe region is passed to the functions as a numpy array of  numpy arrays describing each individual unsafe region. The transition map $f$, represented by \verb|f|, written as a SymPy expression\footnote{\href{https://docs.sympy.org/latest/tutorials/intro-tutorial/basic_operations.html}{https://docs.sympy.org/latest/tutorials/intro-tutorial/basic\_operations.html}} for each dimension using states \texttt{x1,x2,...} and noise parameters \texttt{varsigma1,varsigma2,...} (blue-5). The time horizon $\mathcal{T}$, noted as \verb|t| (blue-7). The distribution of the noise, \verb$NoiseType$, can be specified as either \verb$"normal"$, \verb$"exponential"$, or \verb$"uniform"$ (blue-6). 

Users may also specify \emph{optional} parameters, with default values provided in Listing~\ref{lst:dt-SS}. These include the degree of the Lagrangian multipliers ${l_i}(x),{l_u}(x),l(x)$: \verb$l_degree$ (red-1), which, if not specified (i.e., set to  \verb|None|), will default to the same value as \verb$b_degree$; the type of solver: \verb$solver$ (red-5), that can be either set to \verb$"mosek"$~\cite{mosek} or \verb$"cvxopt"$~\cite{cvxopt}. The confidence level $\phi$ in \eqref{confidence} can be optimized using \verb|optimize| (red-2), if set to \verb|True|. In this case, due to having a bilinearity between $\gamma$ and $\lambda$ in \eqref{confidence}, the user is required to provide one $\lambda$: \verb|lam|, \emph{e.g.,} select $\lambda = 1$ (red-3).
The tool will then optimize for the other decision variables including $\gamma$ and $c$ to provide the highest confidence level $\phi$. Alternatively, the user can select a minimum confidence level $\phi$ (red-4) using \verb|confidence| they desire, so that \textsf{PRoTECT} attempts to search for a BC satisfying that confidence level. The parameters for the distributions should be specified as follows (blue-6): for normal distributions, the mean $\mu$ can be set using \verb|mean|, and the diagonal covariance matrix $\sigma$ can be provided using \verb|sigma|. For exponential distributions, the rate parameter for each dimension can be set using \verb|rate|. For uniform distributions, the boundaries for each dimension can be set using \verb|a| and \verb|b|. We provide two functions for dt-SS (red-6): the first \verb|dt_SS| finds a barrier for a single degree, and the second \verb|parallel_dt_SS| runs the first function in parallel for all barrier degrees up to the maximum barrier degree specified (also called \verb|b_degree|).

\begin{lstlisting}[language=Python,label=lst:dt-SS,caption=dt-SS functions.]
	dt_SS(b_degree, dim, L_initial, U_initial, L_unsafe, U_unsafe, L_space, U_space, x, varsigma, f, t, l_degree=None, NoiseType="normal", optimize=False, solver="mosek", confidence=None, gam=None, lam=None, c_val=None, mean=None, sigma=None, rate=None, a=None, b=None)
 parallel_dt_SS(b_degree, dim, L_initial, U_initial, L_unsafe, U_unsafe, L_space, U_space, x, varsigma, f, t, l_degree=None, NoiseType="normal", optimize=False, solver="mosek", confidence=None, gam=None, lam=None, c_val=None, mean=None, sigma=None, rate=None, a=None, b=None)
\end{lstlisting}

\section{Discrete-Time Deterministic Systems}
Discrete-time deterministic systems (dt-DS) are characterized via $\Sigma_d : x(k+1) = f(x(k)),$
where $f:X\rightarrow X$ is the transition map describing the state evolution of dt-DS. A BC $\barrier:X\rightarrow\mathbb{R}$ for dt-DS can be formulated analogously to Definition~\ref{def:dt-SS-BC}, adhering to the same conditions~\eqref{gamma-condition}-\eqref{lambda-condition}. However, modification is applied to condition~\eqref{dt-SS-condition} while removing the expected value and setting $c = 0$ as
	\begin{equation}
		\label{dt-DS-condition}
		\barrier(f(x)) \leq \barrier(x), \quad\quad\forall x\in X.
	\end{equation}
by applying Lemma~\ref{SOS}, we recast the BC's conditions as SOS polynomials~\eqref{lagrangian multiplier1} and~\eqref{lagrangian multiplier2} as before, and introduce a third SOS polynomial as
\begin{equation}
	\label{lagrangian multiplier dt-DS}
	-\barrier(f(x)) + \barrier(x) - l(x){^\top}g(x).
\end{equation}
Given that the underlying system is deterministic, finding a BC offers a safety guarantee over an \emph{infinite} time horizon~\cite{prajna2004stochastic}.

\subsection{\textsf{PRoTECT} Implementation for dt-DS}
The user is required to input necessary (and optional) parameters as outlined in Subsection~\ref{parameter}, excluding those parameters relevant to stochasticity (\emph{e.g.,} time horizon, constant $c$, noise distribution, and confidence level). Optionally, the user can specify the level sets $\gamma$ using \verb|gam| or $\lambda$ using \verb|lam|. It is important to note that optimization for the level sets $\gamma$ and $\lambda$ is not performed, as any feasible solution with $\lambda > \gamma$ ensures a safety guarantee over an infinite time horizon. Similarly, we provide two functions \verb|dt_DS| and \verb|parallel_dt_DS| for the serial and parallel execution.

\begin{lstlisting}[language=Python,label=lst:dt-DS,caption=dt-DS functions.]
	dt_DS(b_degree,dim, L_initial, U_initial, L_unsafe, U_unsafe, L_space, U_space, x, f, l_degree=None, solver="mosek",gam=None,lam=None)
 parallel_dt_DS(b_degree,dim, L_initial, U_initial, L_unsafe, U_unsafe, L_space, U_space, x, f, l_degree=None, solver="mosek",gam=None,lam=None)
\end{lstlisting}

\section{Continuous-Time Stochastic System}
\label{sec:ct-SS}
Here, we define the notion of barrier certificates for continuous-time stochastic systems (ct-SS). In particular, a ct-SS is defined by a tuple $\Sigma_c^{{\delta}} = (X,f,\delta,\rho)$,
where $X\subseteq \reals^n$ is the state set, $f\!:X \rightarrow X$ is the drift term,  $\delta: \mathbb R^n \rightarrow \mathbb R^{n\times \textsf b}$ is the diffusion term, and $\rho: \mathbb R^n \rightarrow \mathbb R^{n\times \textsf r}$ is the reset term.  A ct-SS $\Sigma_c^{{\delta}}$ satisfies
\begin{equation*}\label{eq:ct-SS:dynamics}
	\Sigma_c^{{\delta}}\!:\mathbf{d}x(t) = f(x(t))\mathbf{d}t + \delta(x(t))\mathbf{d}\mathbb{W}_t + \rho(x(t))\mathbf{d}\mathbb{P}_t,
\end{equation*}
where $\mathbb{W}_t$ and $\mathbb{P}_t$ are \emph{Brownian motions} and \emph{Poisson processes}. We consider the rate of Poisson processes $\mathbb{P}_t^z$ as $\omega_z$ for any $z\in[1,\ldots,r]$.  We assume that all drift, diffusion, and reset terms are  \emph{globally Lipschitz continuous} to ensure existence, uniqueness, and strong Markov property of the solution process~\cite{oksendal2005applied}.

A \emph{twice differentiable} BC $\barrier:X\rightarrow\mathbb{R}^{+}_{0}$ for ct-SS is defined similarly to Definition~\ref{def:dt-SS-BC}, following the same conditions~\eqref{gamma-condition}-\eqref{lambda-condition}, with a modification to condition~\eqref{dt-SS-condition} as
	\begin{equation}
		\label{ct-SS-condition}
		\mathcal{L}\barrier(x) \leq c, \quad\quad\forall x\in X,
	\end{equation}
	where $\mathcal{L}\barrier$ is the \emph{infinitesimal generator} of the stochastic process acting on the function $\barrier:X\rightarrow \reals^{+}_0$, defined as
	\begin{equation*}
		\mathcal{L}\barrier(x) = \partial_x\barrier(x)f(x) + \frac{1}{2}\mathbf{Tr}(\delta(x)\delta(x){^\top}\partial_{x,x}\barrier(x)) + \sum_{j=1}^{r}\omega_j(\barrier(x+\rho(x)\mathbf{e}^r_j)-\barrier(x)),
	\end{equation*}
	where $\mathbf{e}^r_j$ denotes an $r$-dimensional vector with $1$ on the $j$-th entry and $0$ elsewhere.

Applying Lemma~\ref{SOS}, we recast the BC conditions as SOS polynomials~\eqref{lagrangian multiplier1} and~\eqref{lagrangian multiplier2} as before, and introduce a third SOS polynomial:

\begin{equation}\label{lagrangian multiplier ct-SS}
	-\mathcal{L}\barrier(x) - l(x){^\top}g(x) + c.
\end{equation}

For ct-SS, one can adapt Lemma~\ref{Lemma1} to offer the same formulation of confidence level $\phi$ in~\eqref{confidence}, using the constructed level sets $\gamma,\lambda$, and the constant $c$ in~\eqref{lagrangian multiplier ct-SS}.

\subsection{\textsf{PRoTECT} Implementation for ct-SS}
The user is asked to enter necessary (and optional) parameters as detailed in Subsection~\ref{parameter}. Additionally, via the corresponding GUI, users must provide the diffusion term $\delta$ using \verb|delta|
for Brownian motion, the reset term $\rho$ using \verb|rho| 
for Poisson process, and the Poisson process rate $\omega$ using \verb|p_rate|. For cases lacking either Brownian motion or Poisson processes, the corresponding parameter should be set to zero. The confidence level $\phi$ can also be optimized if \verb|optimize| is set to \verb|True|. We provide functions \verb|ct_SS| and \verb|parallel_ct_SS| for the serial and parallel execution. 

\begin{lstlisting}[language=Python,label=lst:ct-SS,caption=ct-SS functions.]
	ct_SS(b_degree, dim, L_initial, U_initial, L_unsafe, U_unsafe, L_space, U_space, x, f, t, l_degree=None, delta=None, rho=None, p_rate=None, optimize=False, solver="mosek", confidence=None, gam=None, lam=None, c_val=None)
 parallel_ct_SS(b_degree, dim, L_initial, U_initial, L_unsafe, U_unsafe, L_space, U_space, x, f, t, l_degree=None, delta=None, rho=None, p_rate=None, optimize=False, solver="mosek", confidence=None, gam=None, lam=None, c_val=None)
\end{lstlisting}

\section{Continuous-Time Deterministic Systems}

Continuous-time deterministic  systems (ct-DS), as the last class of models, can be described by $\Sigma_c : \dot{x} = f(x(t)),\quad t\in\reals^{+}_0$
where $f:X\rightarrow X$ is the vector field. A \emph{differentiable} BC $\barrier:X\rightarrow\mathbb{R}$ for ct-DS is defined akin to Definition~\ref{def:dt-SS-BC}, maintaining conditions~\eqref{gamma-condition}-\eqref{lambda-condition}, with an adjustment to condition~\eqref{dt-SS-condition} with $c = 0$ as
	\begin{equation}
		\label{ct-DS-condition}
		\text{L}_f\barrier(x) \leq 0, \quad\quad\forall x\in X,
	\end{equation}
	where $\text{L}_f\barrier$ is the \emph{Lie derivative} of $\barrier:X\rightarrow\reals$ with respect to vector field $f$, defined as $ \text{L}_f\barrier(x) = \partial_x\barrier(x)f(x).$ Utilizing Lemma~\ref{SOS}, we reformulate the BC into SOS polynomials~\eqref{lagrangian multiplier1},\eqref{lagrangian multiplier2}, while introducing a third SOS polynomial:
\begin{equation}
	\label{lagrangian multiplier ct-DS}
	-{\text{L}}_f\barrier(x) - l(x){^\top}g(x).
\end{equation}
Since the underlying system is deterministic, finding a BC provides a safety assurance over an \emph{infinite} time horizon~\cite{prajna2004stochastic}.

\subsection{\textsf{PRoTECT} Implementation for ct-DS}
The user is expected to input necessary (and optional) parameters as detailed in Subsection~\ref{parameter}, omitting parameters pertinent to stochasticity (\emph{e.g.,} time horizon, constant $c$, and confidence level). Optionally, users can define the level sets $\gamma$ using \verb|gam| or $\lambda$ using \verb|lam|. Optimization for level sets $\gamma$ and $\lambda$ is not conducted, \emph{i.e.,} any feasible solution where $\lambda > \gamma$ ensures a safety guarantee over an infinite time horizon. Functions \verb|ct_DS| and \verb|parallel_ct_DS| are employed for the serial and parallel execution. 

\begin{lstlisting}[language=Python,label=lst:ct-DS,caption=ct-DS functions.]
	ct_DS(b_degree, dim, L_initial, U_initial, L_unsafe, U_unsafe, L_space, U_space, x, f, l_degree=None, solver="mosek", gam=None, lam=None)
 parallel_ct_DS(b_degree, dim, L_initial, U_initial, L_unsafe, U_unsafe, L_space, U_space, x, f, l_degree=None, solver="mosek", gam=None, lam=None)
\end{lstlisting}

\section{Benchmarking and Case Studies}\vspace{-0.2cm}

We highlight specific case studies conducted using PRoTECT, with their performance detailed in the following subsection. Additional case studies are provided in the appendix.\vspace{-0.2cm}

\subsection{Discrete-time Stochastic Systems (dt-SS)}

\subsubsection{Van der Pol Oscillator.}\vspace{-0.2cm}
\begin{figure}[h!]
    \centering
    \includegraphics[width=0.5\textwidth]{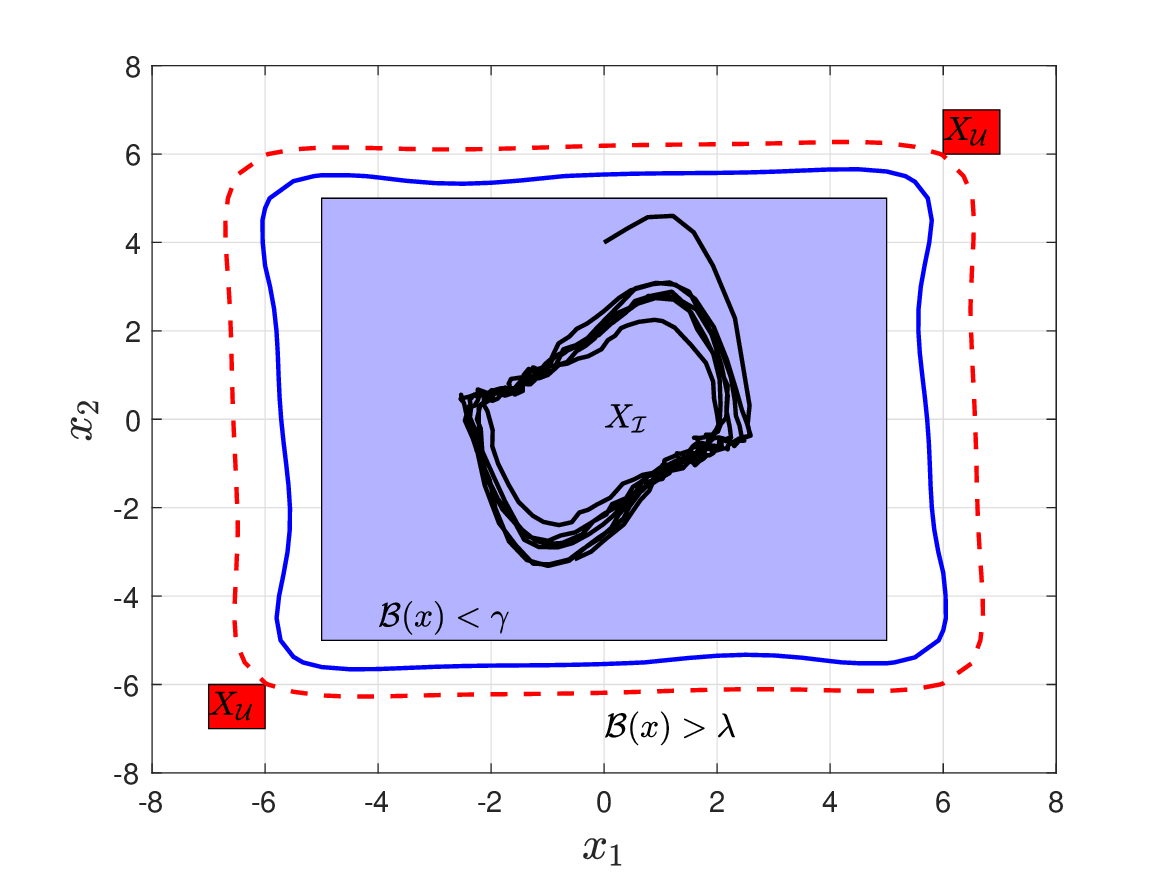}\vspace{-0.2cm}
    \caption{$2$D Van der Pol oscillator system, with initial and unsafe regions $\initial,\unsafe$ marked by blue and red boxes, respectively. Level sets of the barrier certificate are displayed in solid blue and red dashed for $\barrier(x) = \gamma$ and $\barrier(x) = \lambda$, respectively. A trajectory of the system is shown in black with initial point $x_0 = (0,4)$ for a time horizon $\mathcal{T}=5$, with time increments of $0.01$. The uniform noise has bounds [-0.1,0.1] in both dimensions.}\vspace{-0.4cm}
    \label{fig:VDP}
\end{figure}

We consider the stochastic Van der Pol oscillator benchmark from the ARCH competition for stochastic models~\cite{abate2020arch}, with the following dynamics:
\begin{equation*}
    \Sigma_d^\varsigma\!:\begin{cases}
        x_1(k+1) = x_1(k) + \tau x_2(k) + \varsigma_1(k),\\
        x_2(k+1) = x_2(k) + \tau(-x_1(k) + (1-x_1(k)^2)x_2(k)) + \varsigma_2(k),
    \end{cases}
\end{equation*}
where $\tau=0.1$ is the sampling time, and $\varsigma_1$, $\varsigma_2$ are additive noises with uniform distributions with compact supports $[-0.02,0.02]\times[-0.02,0.02]$. We consider $X = [-7,7]^2$, $\initial=[-5,5]^2$ and $\unsafe=[-6,-7]\cup[6,7]$. Fig.~\ref{fig:VDP} shows level sets of $\gamma=400$ and $\lambda=1000$ for a constructed BC.\vspace{-0.2cm}

\subsection{Discrete-time Deterministic Systems (dt-DS)}
\subsubsection{DC Motor.}
\begin{figure}[h!]
    \centering
    \includegraphics[width=0.5\textwidth]{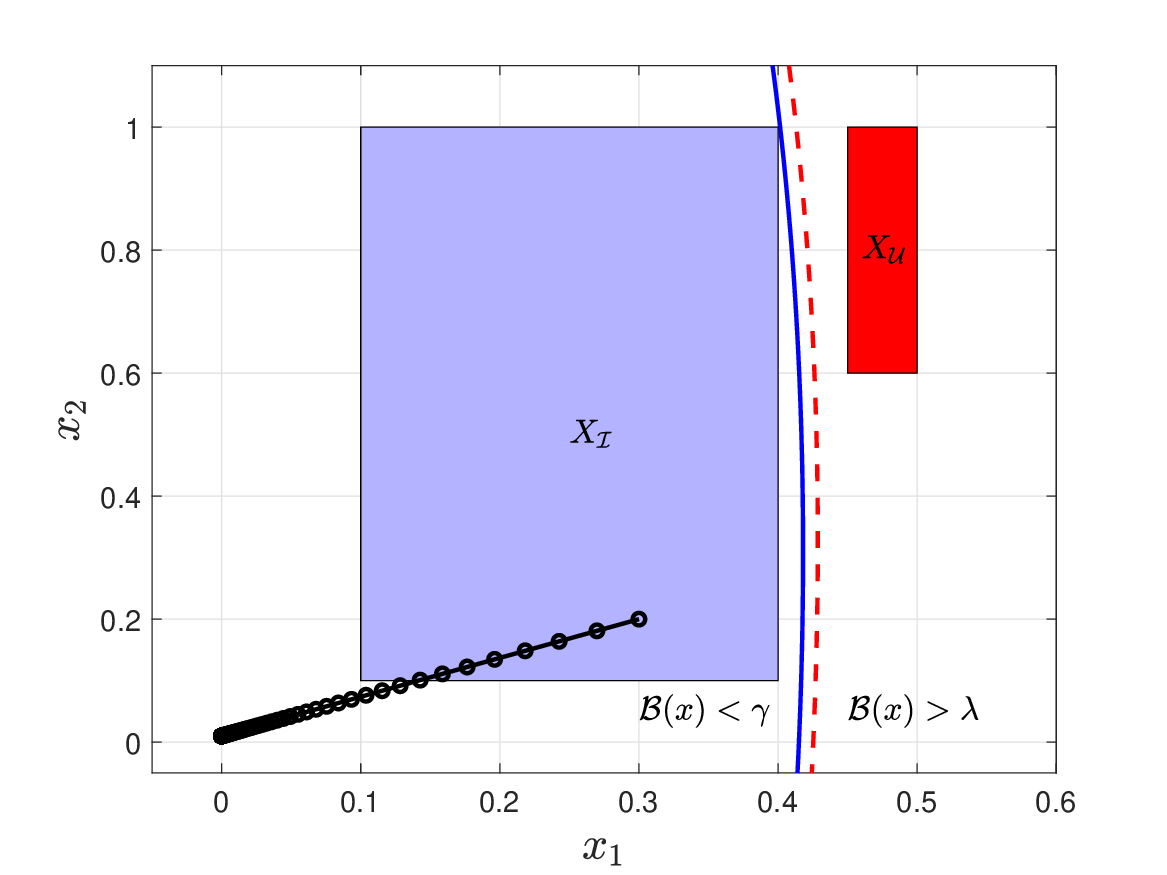}\vspace{-0.2cm}
    \caption{$2$D DC Motor system, with initial and unsafe regions $\initial,\unsafe$ marked by blue and red boxes, respectively. Level sets of the barrier certificate are displayed in solid blue and red dashed for $\barrier(x) = \gamma$ and $\barrier(x) = \lambda$, respectively. A trajectory of the system is shown in black with initial point $x_0 = (0.3,0.2)$ for a time horizon of 100 steps.}\vspace{-0.4cm}
    \label{fig:DC_Motor}
\end{figure}

We consider the following DC motor case study~\cite{adewuyi2013dc}
\begin{equation*}
    \Sigma_d:\begin{cases}
        x_1(k+1) = x_1(k) + \tau(-\frac{R}{L}x_1(k) - \frac{k_{dc}}{L}x_2(k)),\\
        x_2(k+1) = x_2(k) + \tau(\frac{k_{dc}}{J}x_1(k) - \frac{b}{J}x_2(k)),
    \end{cases}
\end{equation*}
where $x_1$ is the armature current, $x_2$ is the rotational speed of the shaft, $\tau$ is the sampling time $0.01$. $R=1,L=0.01,J=0.01,b=1,k_{dc}=0.01$ are, respectively, the electrical resistance, the electrical inductance, the moment of inertia of the rotor, the motor torque and the back electromotive force. We consider regions of interest $X=[0.1,0.5]\times[0.1,1]$, $\initial=[0.1,0.4]\times[0.1,1]$, and $\unsafe=[0.45,0.5]\times[0.6,1]$. Fig.~\ref{fig:DC_Motor} demonstrates a feasible BC with $\gamma=1.19$ and $\lambda=1.20$.

\subsection{Continuous-time Stochastic Systems (ct-SS)}
\subsubsection{Nonlinear System.}
\begin{figure}[h!]
    \centering
    \includegraphics[width=0.5\textwidth]{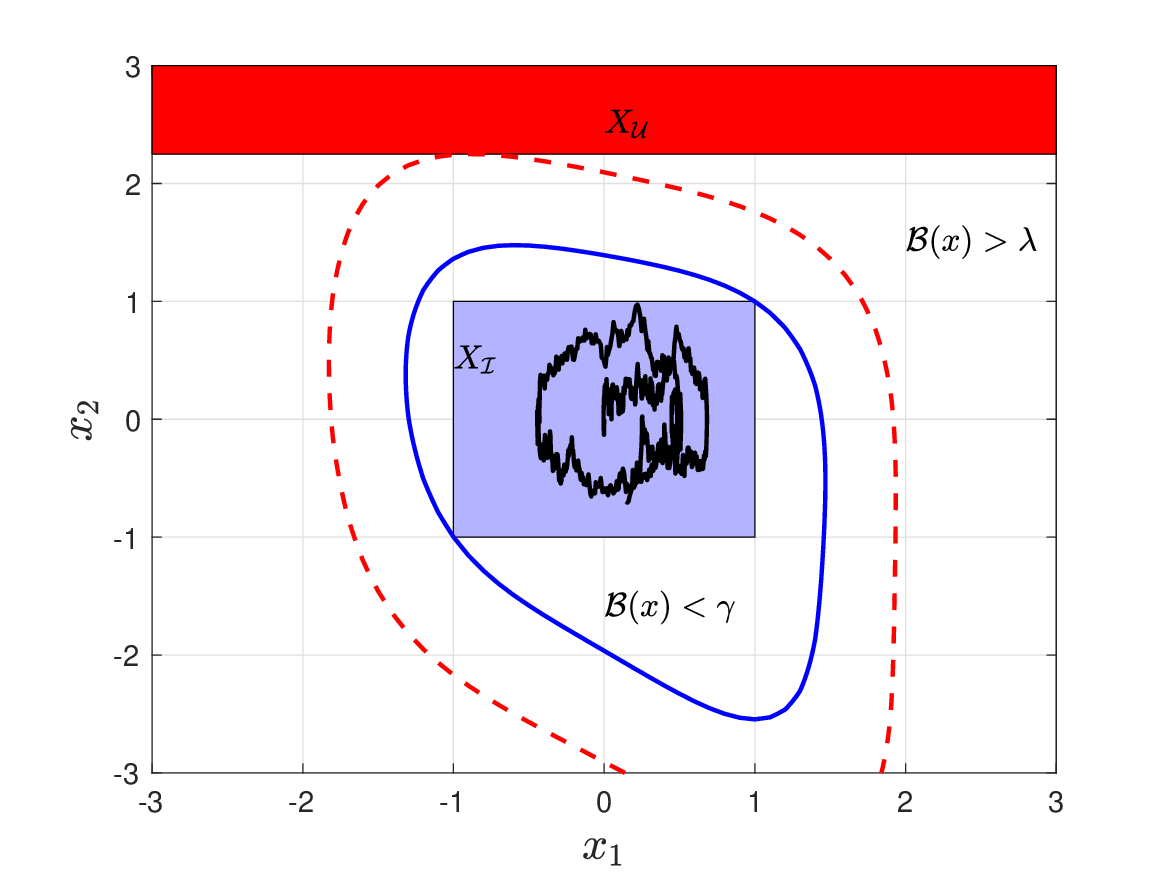}\vspace{-0.2cm}
    \caption{$2$D nonlinear system, with initial and unsafe regions $\initial,\unsafe$ marked by blue and red boxes, respectively. Level sets of the barrier certificate are displayed in solid blue and red dashed for $\barrier(x) = \gamma$ and $\barrier(x) = \lambda$, respectively. A trajectory of the system is shown in black with initial point $x_0 = (0,0)$ for a time horizon $\mathcal{T}=10$ and time increments $0.01$.}\vspace{-0.4cm}
    \label{fig:Nonlinear_System}
\end{figure}

We consider the following nonlinear system~\cite{prajna2004stochastic}
\begin{equation*}
    \Sigma_c^{\delta}:\begin{cases}
    \mathbf{d}{x}_1(t) = x_2(t)\mathbf{d}t, \\
    \mathbf{d}{x}_2(t) = (-x_1(t) -x_2(t) - 0.5x_1^3(t))\mathbf{d}t + \delta\mathbf{d}\mathbb{W}_t,

    \end{cases}
\end{equation*}
where the diffusion term $\delta$ is $0.5$. Regions of interest are $X=[-3,3]^2$, $\initial=[-1,1]^2$, and $\unsafe=[-3,3]\times[2.25,3]$. Fig.~\ref{fig:Nonlinear_System} demonstrates a feasible BC with $\gamma=3.33$ and $\lambda=10$.

\subsection{Continuous-time Deterministic Systems (ct-DS)}
\subsection{Jet Engine Model.}
\begin{figure}[h!]
    \centering
    \includegraphics[width=0.5\textwidth]{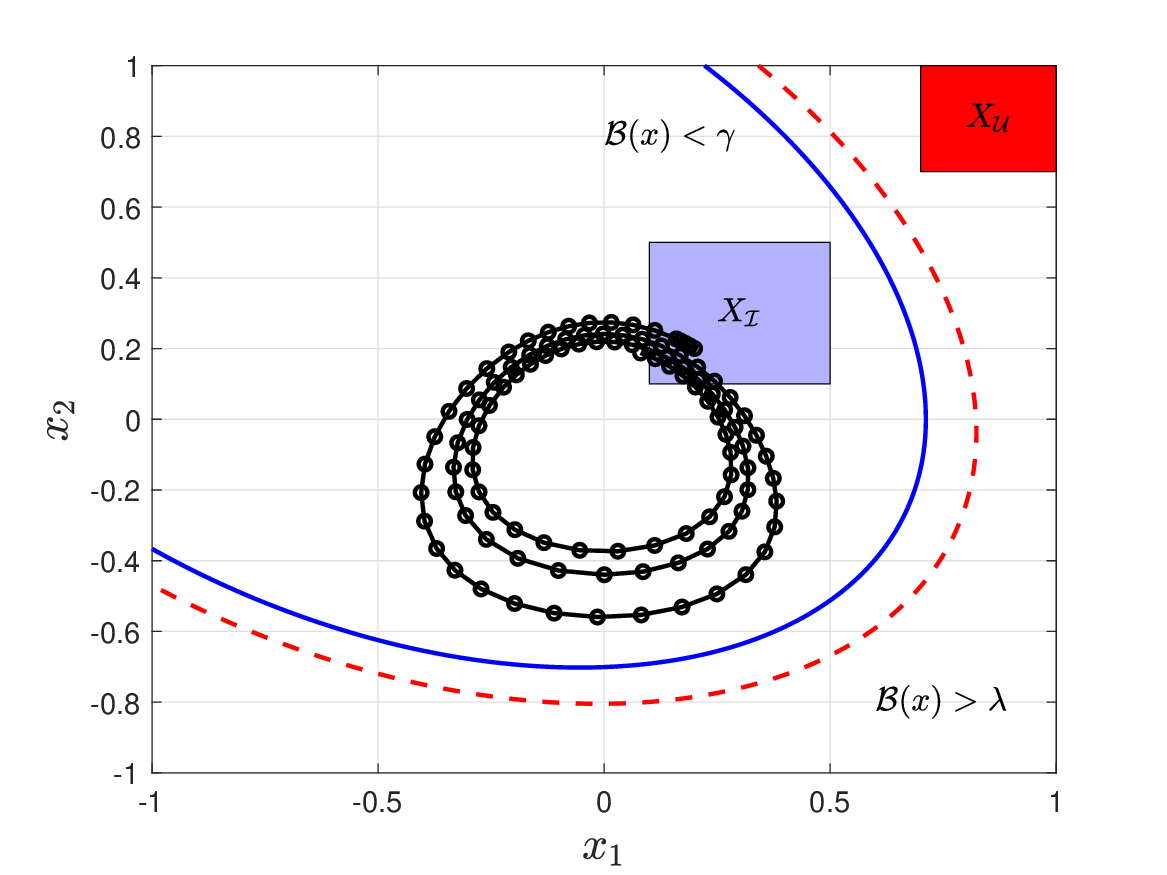}\vspace{-0.2cm}
    \caption{$2$D Jet Engine system, with initial and unsafe regions $\initial,\unsafe$ marked by blue and red boxes, respectively. Level sets of the barrier certificate are displayed in solid blue and red dashed for $\barrier(x) = \gamma$ and $\barrier(x) = \lambda$, respectively. A trajectory of the system is shown in black with initial point $x_0 = (0.2,0.2)$ for a time horizon $\mathcal{T}= 20$.}\vspace{-0.4cm}
    \label{fig:Jet_Engine}
\end{figure}

We consider the nonlinear Moore-Greitzer jet engine model~\cite{krstic1995lean}:
\begin{equation*}
    \Sigma_c: \begin{cases}
        \dot{x}_1(t) = -x_2(t)-1.5x_1^2(t)-0.5x_1^3(t),\\
        \dot{x}_2(t) = x_1(t),
    \end{cases}
\end{equation*}
where $x_1=\Phi-1$, $x_2=\Psi - \Lambda -2$, with $\Phi,\Psi,\Lambda$ being, respectively, the mass flow, pressure rise, and a constant. We consider $X = [0.1,1]^2$, $\initial=[0.1,0.5]^2$, and $\unsafe=[0.7,1]^2$. Fig.~\ref{fig:Jet_Engine} shows level sets $\gamma = 1.41$ and $\lambda = 1.53$ of a feasible BC.

\subsection{Benchmarking}\vspace{-0.2cm}
\begin{table}[h!]
    \centering
    \caption{Efficiency comparison of BC construction for deterministic systems: \textsf{PRoTECT} vs. \textsf{FOSSIL}. All case studies were run on the same desktop computer (Intel i9-12900), and \textsf{PRoTECT} cases were run exclusively for the degree $2$ barrier.}
    \begin{tabular}{c|c|c|c|c|c|c|c}
        \multicolumn{8}{c}{\textsf{One Shot Deterministic Systems}} \\
         &  &  & & & & \multicolumn{1}{|c|}{\textsf{PRoTECT}} & \textsf{FOSSIL} \\
         \hline
          & $n$ & system & \verb|b_degree| &  $\gamma$ & $\lambda$ & (sec) & (sec) \\
         \hline
        $1$D system & 1 & dt-DS &2& {3.90} & {4.04} & {0.09} & {0.20} \\
        DC Motor~\cite{adewuyi2013dc} & 2 & dt-DS&2 & 1.19 & 1.20 & 0.20 & 0.22 \\
        barr$_2$room$_{DT}$ & 2 & dt-DS&2 & 18.0 & 19.5 & 0.17 & 0.44 \\
        Jet Engine~\cite{krstic1995lean} & 2 & ct-DS&2 & 1.41 & 1.53 & 0.18 & 0.28 \\
        hi-ord$_4$ & 4 & ct-DS&2 & 14.5 & 15.0 & 1.30 & 27.5 \\
        hi-ord$_6$-1 & 6 & ct-DS&2 & 32.5 & 33.9 & 17.0 & 11.6 \\
        hi-ord$_{6}$-2 & 6 & ct-DS & 2 & 23.73 & 23.74 & 18.9 & 652 \\
        hi-ord$_8$-1 & 8 & ct-DS&2 & 45.7 & 57.2 & 172 & 21.3 \\
        hi-ord$_{8}$-2 & 8 & ct-DS & 2 & 5.78 & 8.77 & 175 & 569 \\
        \hline
    \end{tabular}
    \label{tab:deterministic_table}
\end{table}

\begin{table}[h!]
    \centering
    \caption{Efficiency evaluation in BC construction for \emph{stochastic systems} via \textsf{PRoTECT}. We present case studies for the barrier degrees 4, optimizing for the barrier with the highest confidence. Parameters $\gamma$ and $c$ are designed via SOS optimization and $\lambda$ is fixed a-priori. VDP denotes the stochastic Van der Pol oscillator. All experiments were conducted on a desktop computer (Intel i9-12900).}
    \begin{tabular}{c|c|c|c|c|c|c|c|c|c}
    & & & & \multicolumn{6}{|c}{\textsf{One Shot Stochastic Systems}} \\
         & $n$ & system & $\mathcal{T}$ & \verb|b_degree| & $\gamma$ & $\lambda$ & $c$ & $\phi$ & time (sec) \\
         \hline
         RoomTemp~\cite{ctSSnejati2021compositional} & 1 & ct-SS & 5 & 4 & $4.8e^{-5}$ & 10 & $9.6e^{-6}$ & 0.99 & 0.16 \\
         RoomTemp~\cite{ctSSnejati2021compositional} & 1 & dt-SS & 5 & 4 & $4.4e^{-5}$ & 10 & $8.9e^{-6}$ & 0.99 & 0.25 \\
         VDP~\cite{abate2020arch} & 2 & dt-SS & 5 & 4 & N/A & 1000 & N/A & N/A & 1.65 \\
         {ex\_lin$_{1}$}~\cite{prajna2004stochastic} & 2 & ct-SS & 5 & 4 & 1.39 & 10 & 0.26 & 0.73 & 0.61 \\
         {ex\_nonlin$_1$}~\cite{prajna2004stochastic} & 2 & ct-SS & 5 & 4 & 3.34 & 10 & 0.53 & 0.40 & 0.61 \\
         TwoTanks~\cite{ramos2007mathematical} & 2 & dt-SS & 5 & 4 & $1.0e^{-6}$ & 10 & $3.4e^{-8}$ & 0.99 & 1.25 \\
         RoomTemp~\cite{salamati2021data} & 3 & dt-SS & 3 & 4 & $1.1e^{-8}$ & 10 & $7.5e^{-9}$ & 0.99 & 29.1 \\
         hi-ord$_4$~\cite{abate2021fossil} & 4 & ct-SS & 3 & 4 & 0.02 & 10 & 0.02 & 0.99 & 68.1 \\
    \end{tabular}
    \label{tab:stochastic_standard}
\end{table}

\begin{table}[h!]
    \centering
    \caption{Efficiency evaluation in BC construction for \emph{stochastic systems} via \textsf{PRoTECT}. We present case studies across three barrier degrees (2, 4, and 6), returning the barrier with the highest confidence. Parameter $\lambda$ is fixed to a certain value and then $\gamma$ and $c$ are designed via  SOS optimization. VDP denotes the stochastic Van der Pol oscillator. All experiments were conducted on a desktop computer (Intel i9-12900).}
    \begin{tabular}{c|c|c|c|c|c|c|c|c|c}
    & & & & \multicolumn{6}{|c}{\textsf{Parallel Stochastic Systems}} \\
         & $n$ & system & $\mathcal{T}$ & \verb|b_degree| & $\gamma$ & $\lambda$ & $c$ & $\phi$ & time (sec) \\
         \hline
         RoomTemp~\cite{ctSSnejati2021compositional} & 1 & ct-SS & 5 & 6 & $1.1e^{-6}$ & 10 & $2.3e^{-7}$ & 0.99 & 0.33 \\
         RoomTemp~\cite{ctSSnejati2021compositional} & 1 & dt-SS & 5 & 6 & $4.4e^{-7}$ & 10 & $1.0e^{-7}$ & 0.99 & 0.53 \\
         VDP~\cite{abate2020arch} & 2 & dt-SS & 5 & 6 & 97.5 & 1000 & 3.53 & 0.88 & 14.3 \\
         {ex\_lin$_{1}$}~\cite{prajna2004stochastic} & 2 & ct-SS & 5 & 6 & 0.34 & 10 & 0.04 & 0.95 & 1.73 \\
         {ex\_nonlin$_1$}~\cite{prajna2004stochastic} & 2 & ct-SS & 5 & 6 & 1.84 & 10 & 0.2 & 0.72 & 1.81 \\
         TwoTanks~\cite{ramos2007mathematical} & 2 & dt-SS & 5 & 4 & $1.0e^{-6}$ & 10 & $3.4e^{-8}$ & 0.99 & 5.14 \\
         RoomTemp~\cite{salamati2021data} & 3 & dt-SS & 3 & 4 & $1.1e^{-8}$ & 10 & $7.5e^{-9}$ & 0.99 & 1501 \\
         hi-ord$_4$~\cite{abate2021fossil} & 4 & ct-SS & 3 & 6 & $1.8e^{-3}$ & 10 & $1.2e^{-3}$ & 0.99 & 1308 \\
    \end{tabular}
    \label{tab:stochastic_optimized}
\end{table}

Table~\ref{tab:deterministic_table} compares the efficiency of \textsf{PRoTECT} and \textsf{FOSSIL}~\cite{abate2021fossil} in finding barrier certificates for deterministic systems, both dt-DS and ct-DS. \textsf{PRoTECT} was operated in a \emph{one-shot mode}, searching for a barrier certificate without parallelism. Unless referenced, the case studies are taken from \textsf{FOSSIL}'s own benchmarks. All examples in this table, including the \textsf{FOSSIL} configuration files, can be located in the folder \texttt{/ex/benchmarks-deterministic}. Notably, \textsf{PRoTECT} demonstrates always better time efficiency for case studies up to four dimensions, while \textsf{FOSSIL} may perform better in cases with six and eight dimensions. This can be attributed to the increase in Lagrangian multipliers in \textsf{PRoTECT}, which also necessitates designing more decision variables when searching for higher-degree BCs. However, as demonstrated in the \textsf{hi-ord$_6$-2} and \textsf{hi-ord$_8$-2} cases, \textsf{FOSSIL} performs significantly worse than \textsf{PRoTECT}, which exhibits superior time efficiency in solving these problems. This is primarily because \textsf{FOSSIL} relies on a neural network approach to solve the problem, and the tool's runtime can vary depending on the network's structure and the corresponding parameters involved, as well as the verification process using SMT solvers. We re-highlight that \textsf{FOSSIL} does not guarantee convergence to a solution.

Table~\ref{tab:stochastic_standard} and Table~\ref{tab:stochastic_optimized} employ \textsf{PRoTECT} for stochastic benchmarks. In the `One Shot' setting with a fixed barrier degree of $4$, as shown in Table~\ref{tab:stochastic_standard}, \textsf{PRoTECT} optimizes all parameters, $\gamma$ and $c$, during the SOS formulation, while $\lambda$ is fixed a priori. Notably, the \textsf{VDP} example cannot find a barrier certificate with degree $4$. We do not provide any comparison with \textsf{FOSSIL} for stochastic systems since, as previously mentioned, \textsf{FOSSIL} cannot handle stochastic systems. In fact, \textsf{PRoTECT} is the first tool in the literature to offer stochastic barrier certificates. Alternatively, the user can set a maximum degree and run computations in parallel up to this degree, returning the barrier certificate with the highest confidence. We call this the ``Parallel'' setting, as shown in Table~\ref{tab:stochastic_optimized}, where $\lambda$ is once again fixed a-priori. We remind the reader that, unlike the deterministic setting, \textsf{PRoTECT} awaits barriers of all degrees to terminate, returning the one with the \emph{highest confidence}. Consequently, stochastic case studies typically require notably longer completion times compared to the ``One Shot'' setting or parallelism for deterministic systems (discussed next), but at the gain of offering a higher confidence, \emph{e.g.} example \textsf{ex\_nonlin$_1$} has a confidence improvement of $32\%$. This is a trade-off the user should navigate based on their particular setting.

\begin{remark}
In Table~\ref{tab:stochastic_standard}, the VDP case does not yield an optimized solution for $\lambda = 1000$. However, by setting the minimum confidence level to $\phi = 0.8$ (as shown in red-4 in Figure~\ref{fig:GUI}), feasible values for $\lambda$, $\gamma$, and $c$ can be identified with confidence exceeding the specified threshold. These results are omitted from Table~\ref{tab:stochastic_standard} to maintain consistency in the comparison.
\end{remark}

Finally, for completeness, we discuss the performance of \textsf{PRoTECT} running in parallel for the deterministic case studies, similar to the stochastic ones. \textsf{PRoTECT} simultaneously searches for a valid BC of varying degrees, such as 2, 4, and 6. Once a valid BC is found, the other threads are terminated. The parallelism introduces minimal overhead when computing multiple results, and we found that running the degrees in parallel is 19-27\% faster than computing the same degrees (2, 4, 6) sequentially. The cases in Table~\ref{tab:deterministic_table} all have valid degree 2 barrier certificates, and we omit the parallelism results due to space constraints.

\section{Conclusion}

This work introduced \textsf{PRoTECT}, a pioneer software tool utilizing \emph{SOS optimization} to explore polynomial-type BCs for verifying safety properties across \emph{four classes of dynamical systems}: dt-SS, dt-DS, ct-SS, and ct-DS. In particular, \textsf{PRoTECT} is the first software tool that offer stochastic barrier certificates. The tool is developed in Python and incorporates a user-friendly GUI to enhance its usability.  Additionally, \textsf{PRoTECT} offers \emph{parallelization} to concurrently search for BCs of different degrees, ensuring an efficient construction. In the future, \textsf{PRoTECT} will be expanded to incorporate reachability and reach-while-avoid specifications, along with designing controllers using SOS optimization techniques.

\bibliographystyle{splncs04}
\bibliography{ref}

\newpage 

\appendix

\section{Case Study Descriptions and Simulation Results}

\subsection{Discrete-time Stochastic Systems (dt-SS)}

\subsubsection{Room Temperature.}
We consider a basic $1$-dimensional room temperature system, based on~\cite{ctSSnejati2021compositional}, with dynamics
\begin{equation*}
    \Sigma_d^\varsigma\!:x(k+1) = (1 - \beta - \theta\nu )x(k) + \theta T_h \nu + \beta T_e + R\varsigma,
\end{equation*}
where $x(k)$ is the temperature of the room, $T_h = 45^{\circ}$C is the heater temperature, $T_e = -15^{\circ}$C is the ambient temperature of the room, $\nu = -0.0120155x + 0.8$, $R=0.1$, and $\beta = 0.6$ and $\theta = 0.145$ are conduction factors. The exponential noise $\varsigma$ has a rate of $1$. Regions of interest are $X = [1,50]$, $\initial = [19.5,20]$, $X_{\mathcal U_1} = [1,17]$, $X_{\mathcal U_2}  = [23,50]$.

\subsubsection{Two-Tank System.}

\begin{figure}[h!]
    \centering
    \includegraphics[width=0.5\textwidth]{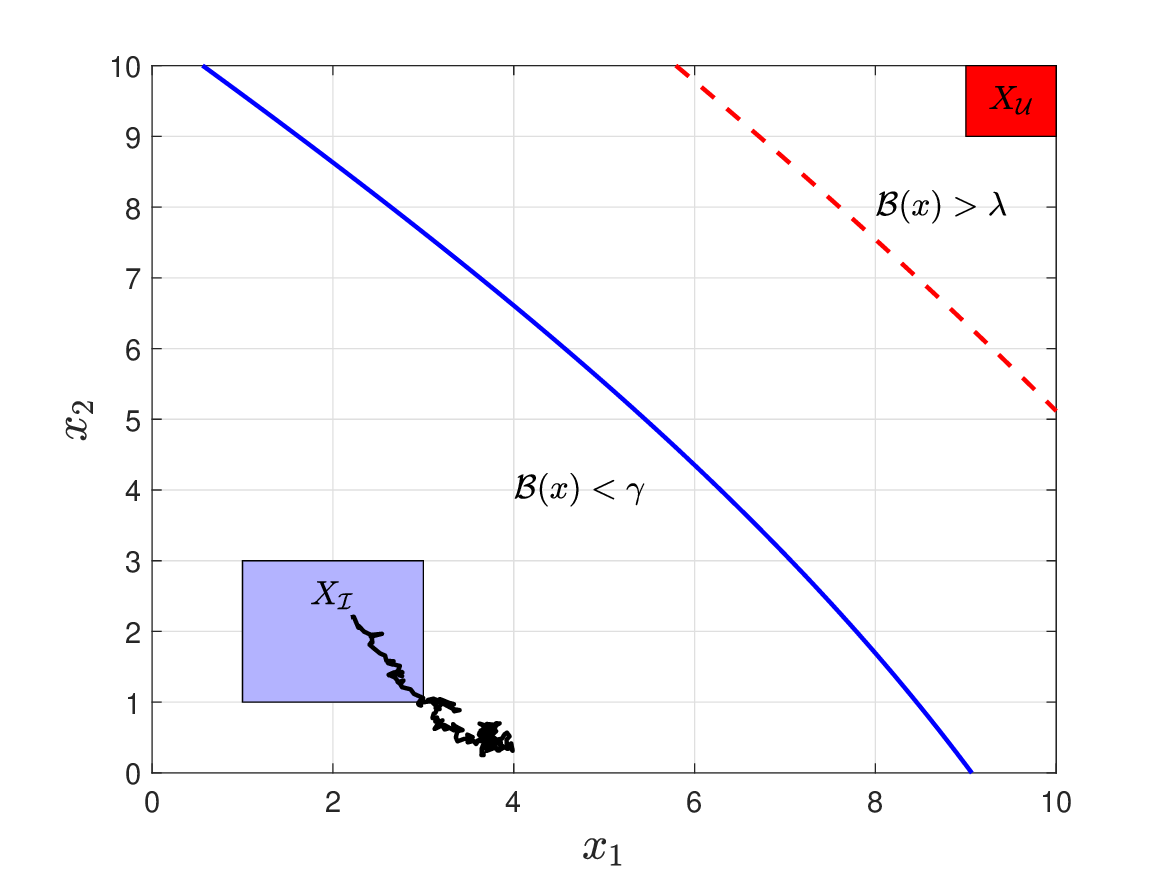}
    \caption{$2$D Two-Tank system, with initial and unsafe regions $\initial,\unsafe$ marked by blue and red boxes, respectively. Level sets of the barrier certificate are displayed in solid blue and red dashed for $\barrier(x) = \gamma$ and $\barrier(x) = \lambda$, respectively. A trajectory of the system is shown in black with initial point $x_0 = (2.2,2.2)$ for $\mathcal{T}=20$, with time increments of $0.01$.}
    \label{fig:TwoTank}
\end{figure}

Consider a two-tank system~\cite{ramos2007mathematical}, characterized by the following difference equations:
\begin{equation*}
    \Sigma_d^\varsigma\!:\begin{cases}
        h_1(k+1) = (1-\tau\frac{\alpha_1}{A_1})h_1(k) + \tau\frac{q_1(k)}{A_1} + \varsigma_1(k),\\
        h_2(k+1) = \tau\frac{\alpha_1}{A_2}h_1(k) + (1-\tau\frac{\alpha_2}{A_2})h_2(k) + \tau\frac{q_o(k)}{A_2} + \varsigma_2(k),
    \end{cases}
\end{equation*}
where $h_1$, $h_2$ are heights of the fluid in two tanks, and $\varsigma_1$, $\varsigma_2$ are Gaussian noises with zero mean and standard deviations of $0.01$. In addition, $\alpha_i$ and $A_i$ are the valve coefficient and area of  tank $i$, and $q_1$ and $q_o$ are the inflow and outflow rate of tank $1$ and $2$, respectively. Furthermore, $\tau=0.1$, $\frac{\alpha_1}{A_1}=1$,$\frac{q_1}{A_1}=4.5$, $\frac{\alpha_1}{A_2}=1$, $\frac{\alpha_2}{A_2}=1$ and $\frac{q_o}{A_2}=-3$. Regions of interest are $X=[1,10]\times[1,10]$, $\initial=[1.75,2.25]\times[1.75,2.25]$, $\unsafe=[9,10]\times[9,10]$. Fig.~\ref{fig:TwoTank} shows level sets $\gamma=1$ and $\lambda=10$ of a constructed BC.

\subsubsection{3D Room Temperature.} 
We consider the $3$-dimensional room temperature case study~\cite{lavaei2020amytiss}, with dynamics
\begin{equation*}
    \Sigma_d^\varsigma\!:\begin{cases}
        x_1(k+1) = (1-\tau(\alpha+\alpha_e))x_1(k) + \tau\alpha x_2(k) + \tau\alpha_e T_e + \varsigma_1(k),\\
        x_2(k+1) = (1-\tau(2\alpha+\alpha_e))x_1(k) + \tau\alpha (x_1(k)+x_3(k)) + \tau\alpha_e T_e + \varsigma_2(k),\\
        x_3(k+1) = (1-\tau(\alpha+\alpha_e))x_3(k) + \tau\alpha x_2(k) + \tau\alpha_e T_e + \varsigma_3(k),
    \end{cases}
\end{equation*}
where $x_1,x_2,x_3$ are temperatures of the three rooms, $\varsigma_1,\varsigma_2,\varsigma_3$ are zero-mean Gaussian noises with standard deviation $0.01$, $T_e = 10^{\circ}$C is the ambient temperature, $\alpha_e = 8\times10^-3$ and $\alpha=6.2\times 10^-3$ are heat exchange coefficients, and $\tau=5$ minutes is the sampling time. Regions of interest are $X = [17,30]^3$, $\initial=[17,18]^3$ and $\unsafe=[29,30]^3$. 

\subsection{Discrete-time Deterministic Systems (dt-DS)}

\subsubsection{1D Basic System.}

We consider a simple scalar test case, with dynamics
\begin{equation*}
    \Sigma_d: x(k+1) = x(k) + \tau (15 - x(k) + 0.1\alpha_h (55 - x(k))),
\end{equation*}
where $\tau=5,\alpha_h=3.6\times10^{-3}$ with regions of interest  $X=[-6,6]$, $\initial=[-0.5,0.5]$, and $\unsafe=[-6,-5]$.

\subsubsection{\textsf{FOSSIL} Benchmark - $2$D Room System.}

\begin{figure}[h!]
    \centering
    \includegraphics[width=0.5\textwidth]{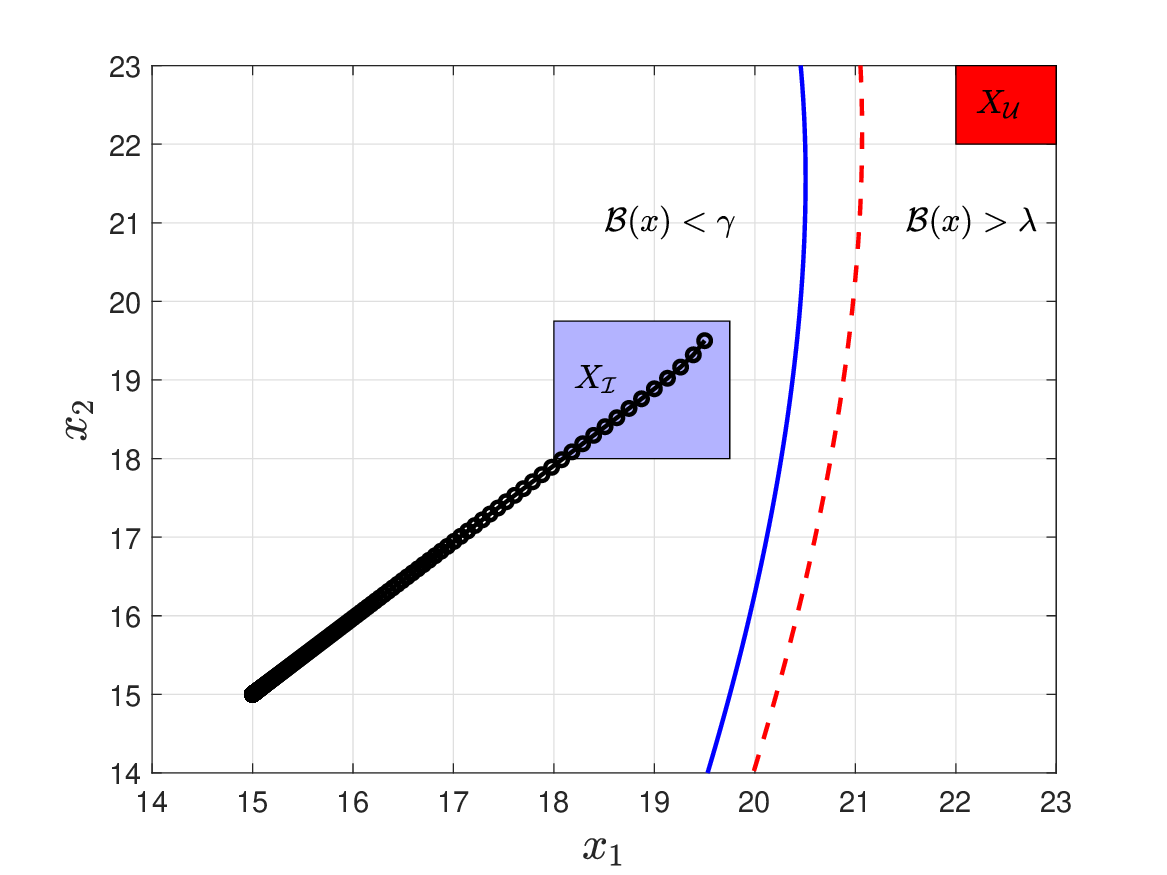}
    \caption{$2$D Room system, with initial and unsafe regions $\initial,\unsafe$ marked by blue and red boxes, respectively. Level sets of the barrier certificate are displayed in solid blue and red dashed for $\barrier(x) = \gamma$ and $\barrier(x) = \lambda$, respectively. A trajectory of the system is shown in black with initial point $x_0 = (19.5,19.5)$ for a time horizon of 1000 steps.}
    \label{fig:TwoRoom}
\end{figure}

We consider the two room system from the \textsf{FOSSIL} benchmarks~\cite{abate2021fossil}, with dynamics:
\begin{equation*}
    \Sigma_d:\begin{cases}
        x_1(k+1) = (1 - \tau(\alpha + \alpha_{e1}))x_1(k) + \tau\alpha x_2(k) + \tau\alpha_{e1} T_e,\\
        x_2(k+1) = (1 - \tau(\alpha + \alpha_{e2}))x_2(k) + \tau\alpha x_1(k) + \tau\alpha_{e2}T_e,
    \end{cases}
\end{equation*}
where the discretization parameter $\tau = 5$, heat exchange constants $\alpha = 5\times 10^{-2}, \alpha_{e1} = 5\times 10^{-3}, \alpha_{e2} = 8\times10^{-3}$, and external temperature $T_e = 15$. We consider regions of interest $X=[18,23]^2$, $\initial=[18,19.75]^2$, and $\unsafe=[22,23]$. Fig.~\ref{fig:TwoRoom} demonstrates a feasible BC with $\gamma=17.9$ and $\lambda=19.2$.

\subsection{Continuous-time Stochastic Systems (ct-SS)}

\subsubsection{Room Temperature.}
We consider a room temperature system with dynamics~\cite{ctSSnejati2021compositional}
\begin{equation*}
    \Sigma_c^\delta\!:\mathbf{d}{x}(t) = ((-2\eta - \beta - \theta\nu) x(t) + \theta T_h \nu + \beta T_e)\mathbf{d}t + \delta\mathbf{d}\mathbb{W}_t + \rho\mathbf{d}\mathbb{P}_t,
\end{equation*}
where $x$ is the temperature of the room, $T_h = 48^{\circ}$C is the heater temperature, $T_e = -15^{\circ}$C is the outside temperature, $\nu = -0.0120155x + 0.7$, and $\eta=0.005$, $\beta = 0.6$ and $\theta = 0.156$ are conduction factors. The system noise consists of both both Brownian with diffusion term $\delta=0.1$ and Poisson process with reset term $\rho=0.1$ and rate $0.1$. Region of interest are $X = [1,50]$, $\initial = [19.5,20]$, and  $X_{\mathcal U_1} = [1,17]$, $X_{\mathcal U_2}  = [23,50]$.

\subsubsection{2D Linear System.}
\begin{figure}[h!]
    \centering
    \includegraphics[width=0.5\textwidth]{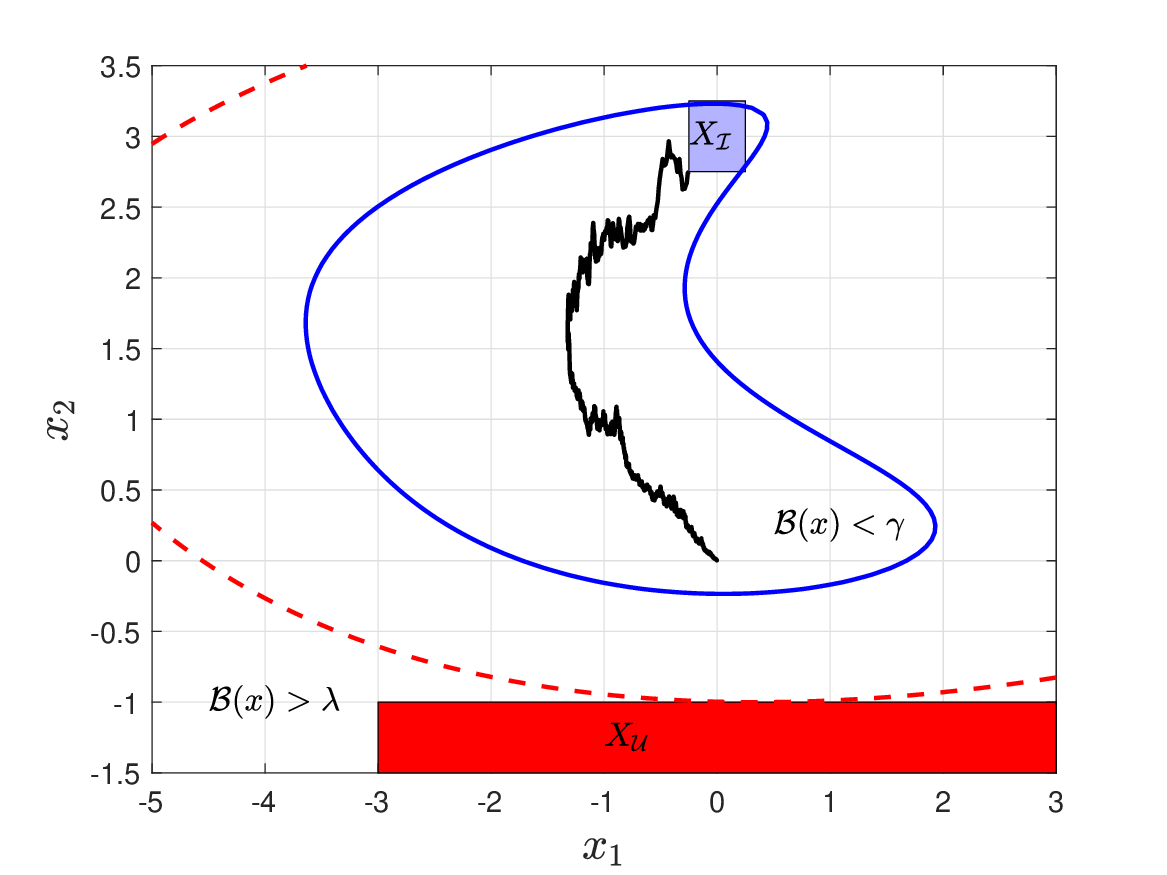}
    \caption{$2$D linear system, with initial and unsafe regions $\initial,\unsafe$ marked by blue and red boxes, respectively. Level sets of the barrier certificate are displayed in solid blue and red dashed for $\barrier(x) = \gamma$ and $\barrier(x) = \lambda$, respectively. A trajectory of the system is shown in black with initial point $x_0 = (-0.25,2.75)$ for a time horizon $\mathcal{T}=5$, and time increments $0.001$.}
    \label{fig:Linear_System}
\end{figure}

We consider the following linear example~\cite{prajna2004stochastic}
\begin{equation*}
    \Sigma_c^{\delta}:\begin{cases}
    \mathbf{d}{x}_1(t) = (-5x_1(t) -4x_2(t))\mathbf{d}t, \\
    \mathbf{d}{x}_2(t) = (-x_1(t)-2x_2(t))\mathbf{d}t + \delta(x_2(t))\mathbf{d}\mathbb{W}_t,
    \end{cases}
\end{equation*}
where the diffusion term $\delta(x_2(t))$ is $0.5x_2(t)$. 
Regions of interest are $X=[-3,3]\times[-1.5,3.5]$, $\initial=[-0.25,0.25]\times[2.75,3.25]$, and $\unsafe=[-3,3]\times[-1.5,-1]$. Fig.~\ref{fig:Linear_System} demonstrates a feasible BC with $\gamma=1.38$ and $\lambda=10$.

\subsubsection{Benchmark - High Order 4.}

We consider the following $4$-dimensional dynamics
\begin{equation*}
    \Sigma_c^{\delta}:\begin{cases}
        \mathbf{d}{x}_1(t) = x_2(t)\mathbf{d}t + \delta_1\mathbf{d}\mathbb{W}_t,\\
        \mathbf{d}{x}_2(t) = x_3(t)\mathbf{d}t + \delta_2\mathbf{d}\mathbb{W}_t,\\
        \mathbf{d}{x}_3(t) = x_4(t)\mathbf{d}t + \delta_3\mathbf{d}\mathbb{W}_t,\\
        \mathbf{d}{x}_4(t) = (-3980x_4(t) - 4180x_3(t) - 2400x_2(t) - 576x_1(t))\mathbf{d}t + \delta_4\mathbf{d}\mathbb{W}_t,
    \end{cases}
\end{equation*}
with diffusion terms $\delta_i = 0.1,$ for all $i\in\{1,\dots,4\}$. The regions of interest are $X=[-2,2]^4$, $\initial=[0.5,1.5]^4$, and $\unsafe=[-2.4,-1.6]^4$.

\subsection{Continuous-time Deterministic Systems (ct-DS)}

\subsubsection{\textsf{FOSSIL} Benchmark - High Order 4 (hi-ord$_4$).}

We consider the following $4$-dimensional benchmark~\cite{abate2021fossil}
\begin{equation*}
    \Sigma_c:\begin{cases}
        \dot{x}_1(t) = x_2(t),\\
        \dot{x}_2(t) = x_3(t),\\
        \dot{x}_3(t) = x_4(t),\\
        \dot{x}_4(t) = -3980x_4(t) - 4180x_3(t) - 2400x_2(t) - 576x_1(t),
    \end{cases}
\end{equation*}
with the state space $X=[-2,2]^4$, initial region $\initial=[0.5,1.5]^4$, and unsafe region $\unsafe=[-2.4,-1.6]^4$.

\subsubsection{\textsf{FOSSIL} benchmark - High Order 6 (hi-ord$_6$-1).}

We consider the following $6$-dimensional benchmark~\cite{abate2021fossil}
\begin{equation*}
    \Sigma_c:\begin{cases}
        \dot{x}_1(t) = x_2(t),\\
        \dot{x}_2(t) = x_3(t),\\
        \dot{x}_3(t) = x_4(t),\\
        \dot{x}_4(t) = x_5(t),\\
        \dot{x}_5(t) = x_6(t),\\
        \dot{x}_6(t) = -800x_6(t) - 2273x_5(t) -3980x_4(t) - 4180x_3(t) - 2400x_2(t) - 576x_1(t),
    \end{cases}
\end{equation*}
with the state space $X=[-2,2]^6$, initial region $\initial=[0.5,1.5]^6$, and unsafe region $\unsafe=[-2.4,-1.6]^6$.

\subsubsection{High Order 6 (hi-ord$_6$-2)}

We consider the following $6$-dimensional example
\begin{equation*}
    \Sigma_c:\begin{cases}
        \dot{x}_1(t) = x_2(t)-100x_3(t),\\
        \dot{x}_2(t) = x_3(t),\\
        \dot{x}_3(t) = x_4(t)-100x_5(t),\\
        \dot{x}_4(t) = x_5(t),\\
        \dot{x}_5(t) = x_6(t)-100x_1(t),\\
        \dot{x}_6(t) = -800x_6(t) - 2273x_5(t) -3980x_4(t) - 4180x_3(t) - 2400x_2(t) - 576x_1(t),
    \end{cases}
\end{equation*}
with the state space $X=[-2,2]^6$, initial region $\initial=[0.5,1.5]^6$, and unsafe region $\unsafe=[-2.4,-1.6]^6$.

\subsubsection{\textsf{FOSSIL} benchmark - High Order 8 (hi-ord$_8$-1).}
We consider the following $8$-dimensional benchmark~\cite{abate2021fossil}
\begin{equation*}
    \Sigma_c:\begin{cases}
        \dot{x}_1(t) = &x_2(t),\\
        \dot{x}_2(t) = &x_3(t),\\
        \dot{x}_3(t) = &x_4(t),\\
        \dot{x}_4(t) = &x_5(t),\\
        \dot{x}_5(t) = &x_6(t),\\
        \dot{x}_6(t) = &x_7(t),\\
        \dot{x}_7(t) = &x_8(t),\\
        \dot{x}_8(t) = &-20x_8(t)-170x_7(t)-800x_6(t) - 2273x_5(t) -3980x_4(t) - 4180x_3(t) \\ &- 2400x_2(t) - 576x_1(t),
    \end{cases}
\end{equation*}
with the state space $X=[-2.2,2.2]^8$, initial region $\initial=[0.9,1.1]^8$, and unsafe region $\unsafe=[-2.2,-1.8]^8$.

\subsubsection{High Order 8 (hi-ord$_8$-2).}
We consider the following $8$-dimensional example
\begin{equation*}
    \Sigma_c:\begin{cases}
        \dot{x}_1(t) = &x_2(t)-50x_3(t),\\
        \dot{x}_2(t) = &x_3(t)-50x_4(t),\\
        \dot{x}_3(t) = &x_4(t)-50x_5(t),\\
        \dot{x}_4(t) = &x_5(t)-50x_6(t),\\
        \dot{x}_5(t) = &x_6(t)-50x_7(t),\\
        \dot{x}_6(t) = &x_7(t)-50x_8(t),\\
        \dot{x}_7(t) = &x_8(t)-50x_1(t),\\
        \dot{x}_8(t) = &-20x_8(t)-170x_7(t)-800x_6(t) - 2273x_5(t) -3980x_4(t) - 4180x_3(t) \\ &- 2400x_2(t) - 576x_1(t),
    \end{cases}
\end{equation*}
with $X=[-2.2,2.2]^8$, $\initial=[0.9,1.1]^8$, and $\unsafe=[-2.2,-1.8]^8$.

\end{document}